\documentstyle[aps,twocolumn]{revtex}

\begin{document}
\title{The Kosterlitz-Thouless and magnetic transition temperatures in layered
magnets with a weak easy-plane anisotropy}
\author{V.Yu.Irkhin$^{*}$ and A.A.Katanin}
\address{620219, Institute of Metal Physics, Ekaterinburg, Russia.}
\maketitle

\begin{abstract}
The two-dimensional (2D) Heisenberg magnet with a weak easy-plane anisotropy
is considered. A renormalization group (RG) analysis in this model is
performed for both quantum and classical cases. A crossover from the
Heisenberg to 2D $XY$ model is discussed. The magnetic transition owing to
the interlayer coupling is considered. Analytical results for the
Kosterlitz-Thouless and Curie (N\'eel) temperatures are derived with account
of two-loop corrections. The results are compared with experimental data,
e.g. on K$_2$CuF$_4,$ and turn out to provide a quantitative description,
unlike the standard one-loop results.
\end{abstract}


The interest in the magnetic properties of layered systems has been recently
greatly revived. It is well known that even weak magnetic anisotropy can
play in such systems an important role. In the present paper we discuss the
case of the easy-plane localized-spin system. The simplest classical
two-dimensional (2D) $XY$ model was studied in detail\cite{Kosterlitz,Jose},
and the relevance of topological (vortex) excitations in thermodynamics was
established. In particular, the Kosterlitz-Thouless transition, connected
with unbinding of the vortex-antivortex pairs, was found. The transition
temperature, where power-law behavior of spin correlation function is
changed by exponential one, is estimated as 
\begin{equation}
T_{KT}=\frac \pi 2JS^2
\end{equation}
where $J>0$ is the exchange integral. In the quantum $XY$ model the
situation is still more complicated, since not only transverse, but also $z$%
-components of spins should be taken into account.

A different situation takes place in both quantum and classical Heisenberg
2D model with a weak easy-plane anisotropy, which is a more physically real
case \cite{Joungh}. A simple expression for Kosterlitz-Thouless temperature
obtained in Ref. \cite{Hikami} reads 
\begin{equation}
T_{KT}=\frac{4\pi JS^2}{\ln [\pi ^2/(1-\eta )]}  \label{TKTs}
\end{equation}
($\eta =J^z/J^{x,y}$ is the anisotropy parameter) and has the same form as
the result for the magnetic ordering point of the easy-axis layered magnet%
\cite{Joungh}. As well as the latter result (see discussion in Refs.\cite
{Our1/N,OurLett,OurRG}), the formula (\ref{TKTs}) is insufficient for a
quantitative description of the experimental data (see Ref. \cite{Hikami}).
Since $T_{KT}\ll JS^2$, one can expect that thermodynamic properties of
these systems are determined by usual spin waves, except for a narrow region
near $T_{KT}$. The situation is reminiscent of the easy-axis layered magnet 
\cite{Our1/N,OurLett,OurRG}, where the topological (domain-wall) excitations
are important only in the vicinity of the Curie (N\'eel) temperature $%
T_C(T_N)$ In such a situation, similar to \cite{OurRG}, the
renormalization-group (RG) analysis can be performed to calculate $T_{KT}$
with higher accuracy. This analysis is the aim of the present paper. Further
on, we consider effects of interlayer coupling, which lead to occurrence of
the true long-range magnetic ordering, and calculate $T_C(T_N)$.

We consider the easy-plane Heisenberg model 
\begin{equation}
H=-\frac 12\sum_{\langle ij\rangle }J_{ij}\left[ S_i^xS_j^x+S_i^yS_j^y+\eta
S_i^zS_j^z\right]  \label{H}
\end{equation}
where $J_{ij}=J\;$($J>0$ in the ferromagnetic (FM) case and $J<0$ in the
antiferromagnetic (AF) case) for the nearest-neighbor sites $i,j$ in the
same plane, and $J_{ij}=\alpha J$ for $i,j$ in different planes, $\eta <1$.
We suppose $1-\eta ,\,\alpha $ $\ll 1.$ Note that the effect of the
single-site anisotropy 
\begin{equation}
H_D=D\sum_i(S_i^z)^2,\;D>0
\end{equation}
is the same as that of exchange anisotropy with $1-\eta =D(1-1/2S)/4|J|$
provided that $D/\,|J|$ $\ll 1.$

The partition function of the model (\ref{H}) can be represented in terms of
a path integral over coherent states (see, e.g., Refs.\cite
{Klauder,ArovasBook}): 
\begin{eqnarray}
{\cal Z} &=&\int D\mbox {\boldmath $\pi $}\delta (\pi ^2-1)\exp (-{\cal L}%
_{dyn}-{\cal L}_{st})  \label{zp} \\
{\cal L}_{\text{dyn}} &=&\text{i}S\int\limits_0^{1/T}d\tau \sum_i{\bf A}(%
\mbox {\boldmath $\pi $}_i)\frac{\partial \mbox {\boldmath $\pi $}_i}{%
\partial \tau }  \nonumber \\
{\cal L}_{\text{st}} &=&\frac 12S^2\int\limits_0^{1/T}d\tau \sum_{\langle
ij\rangle }J_{ij}\left[ \pi _{xi}\pi _{xj}+\pi _{yi}\pi _{yj}+\eta \pi
_{zi}\pi _{zj}\right]  \nonumber
\end{eqnarray}
where $\mbox {\boldmath $\pi $}$ is the unit-length vector, and ${\bf A}(%
\mbox {\boldmath $\pi $})$ is the vector potential of the unit magnetic
monopole, which satisfies the equation ${\bf \nabla }\times {\bf A(}%
\mbox
{\boldmath $\pi $}{\bf )}\cdot \mbox {\boldmath $\pi $}=1.$

In the classical case (i.e., with ${\cal L}_{\text{dyn}}$ being neglected)
we have two types of excitations: the field $\pi _y$ describes the gapless
in-plane excitations, and the field $\pi _z$ describes the out-of-plane
excitations with a gap. Expanding (\ref{zp}) in $\pi _{y,z},$ $x$ being a
(local in AF case) spin quantization axis, to leading order in $1/S$ we have 
\begin{eqnarray}
{\cal L}_{\text{st}} &=&\frac 12S^2\int\limits_0^{1/T}d\tau \sum_{{\bf k}%
}\left[ (J_0-J_{{\bf k}})\pi _{y{\bf k}}\pi _{y,-{\bf k}}\right.  \nonumber
\\
&&\ \ \ \ \ \ \left. +(J_0-\eta J_{{\bf k}})\pi _{z{\bf k+Q}}\pi _{z,-{\bf %
k-Q}}\right]
\end{eqnarray}
where ${\bf Q}$ is the wavevector of magnetic structure. The dynamical part $%
{\cal L}_{\text{dyn}},$ which is present in the quantum case, results in (i)
quantum renormalizations of the Hamiltonian parameters (in AF case only),
which are supposed to be already performed and (ii) the summations over
wavevectors are bounded by $\sqrt{T/JS}$ in FM case or $T/c$ in AF case ($%
c\; $is the quantum-renormalized spin-wave velocity) rather than by the
Brillouin zone boundary (see Ref. \cite{OurRG}).

The interaction of spin waves, which occurs in higher orders in $1/S,$ leads
to temperature renormalizations of the Hamiltonian parameters. Due to the
smallness of anisotropy, large logarithms occur in these renormalizations: $%
\ln |T/(1-\eta )J|$ in the quantum FM case, $\ln [T^2/(1-\eta )J^2]$ in the
quantum AF case, and $\ln [1/(1-\eta )]$ in the classical case. It is
natural to sum up these logarithms within the RG approach. However, there
exists the difficulty owing to that gapless $\pi _y$-excitations are present
too. In the absence of the interlayer coupling, they lead to infrared
divergences in some quantities like the in-plane magnetization. In the
presence of interlayer coupling, another type of logarithms occur, $\ln
(T/\alpha J),$ $\ln (T^2/\alpha J^2)$ or $\ln (1/\alpha ),$ depending on the
case. The situation, where two types of excitations with different
characteristic scales are present, is typical for systems demonstrating a
crossover \cite{Amit}. In our model this is the crossover from the
Heisenberg (almost isotropic) behavior to the $XY$ behavior.

To describe correctly this crossover we include anisotropy in all the
renormalization factors \cite{Amit}. We also introduce the scaling factors
of the field $\mbox {\boldmath $\pi $}$. Because of anisotropic character of
the model, we have two such factors: $Z_{xy}$ and $Z_z{\bf ,}$ so that $\pi
_{xR}/\pi _x=\pi _{yR}/\pi _y=Z_{xy}$ and $\pi _{zR}/\pi _z=Z_z.$ We use the
normalization condition $\Gamma _{zz}^{(2)}(0)=1-\eta $ (which fixes the gap
of $z$-excitations)$\;$instead of the standard one, $d\Gamma
_{zz}^{(2)}(q)/d(q^2)=1,$ $\Gamma _{zz}^{(2)}(q)$ being the two-point vertex
function (inverse Green's function) of the field $\pi _z$. Then we have $%
Z_z\equiv 1.$ For other Hamiltonian parameters we obtain the following
system of RG equations 
\begin{mathletters}
\label{Eq}
\begin{eqnarray}
\mu \frac{d(1/t_\mu )}{d\mu } &=&(1+t_\mu )f(\eta _\mu ,\mu )+{\cal O}(t_\mu
^2)  \label{Eq1} \\
\mu \frac{d\ln Z_{xy}}{d\mu } &=&t_\mu \left[ 1+f(\eta _\mu ,\mu )\right] +%
{\cal O}(t_\mu ^3)  \label{Eq2} \\
\mu \frac{d\ln \eta _\mu }{d\mu } &=&2t_\mu f(\eta _\mu ,\mu )+{\cal O}%
(t_\mu ^2)  \label{Eq3} \\
\mu \frac{d\ln \alpha _\mu }{d\mu } &=&-t_\mu +{\cal O}(t_\mu ^2)
\label{Eq4}
\end{eqnarray}
where $\mu \;$is the scale parameter, $f(\eta _\mu ,\mu )=\eta _\mu \mu
^2/(\eta _\mu \mu ^2+1-\eta ),\;$%
\end{mathletters}
\[
t=\left\{ 
\begin{array}{cc}
T/(2\pi JS^2) & \text{FM} \\ 
T/(2\pi \rho _s) & \text{AF}
\end{array}
\right. 
\]
is the dimensionless temperature, $\rho _s\simeq S(S+0.079)|J|$ being the
spin stiffness \cite{Chakraverty}. First two equations in (\ref{Eq}) are
written down to two-loop order, while last two to one-loop order, which is
sufficient to obtain the final results to the two-loop order accuracy. The
initial scale $\mu _0$ for these equations is 
\[
\mu _0=\left\{ 
\begin{array}{cc}
\sqrt{32} & \text{classical regime (}T\gg |J|S\text{)} \\ 
T/c & \text{quantum regime (}T\ll |J|S\text{), AF} \\ 
\sqrt{T/JS} & \text{quantum regime (}T\ll JS\text{), FM}
\end{array}
\right. ,
\]
for the details see, e.g., Ref. \cite{OurRG}.

The flow of RG parameters is shown schematically in Fig.1 (for comparison,
the easy-axis case $\eta >1$ is depicted too). From the equations (\ref{Eq2}%
) and (\ref{Eq4}) we obtain 
\begin{equation}
\frac 1{t_\mu }=\frac 1t+\frac 12\ln \frac{\eta \mu ^2t_\mu ^2+t^2(1-\eta )}{%
\eta \mu _0^2t_\mu ^2+t^2(1-\eta )}+\ln \frac t{t_\mu }+\Phi (\mu )
\end{equation}
where the function $\Phi (\mu )={\cal O}(t_\mu )$ comes from ${\cal O}$%
-terms in (\ref{Eq}) and corresponds to the contribution of higher-order
loops. For $\mu \gg \sqrt{1-\eta }$ the effective temperature $t_\mu $ is
small (which guarantees that the spin-wave theory works well), so that we
have $\Phi (\mu )\ll 1$ and 
\begin{equation}
\frac 1{t_\mu }=\frac 1t+\ln \frac{\mu t}{\mu _0t_\mu }
\end{equation}
For $\mu \ll \sqrt{1-\eta }$ we obtain 
\begin{equation}
\frac 1{t_\mu }=\frac 1t-\ln \frac{\mu _0}{\sqrt{1-\eta }}+2\ln \frac t{%
t_\mu }+\Phi (\mu )  \label{tr}
\end{equation}
and in this regime $t_\mu $ is $\mu $-dependent only through the function $%
\Phi (\mu )$. The scale $1/\sqrt{1-\eta }$ is just a characteristic scale
for the crossover from the Heisenberg to $XY$ behavior and (\ref{tr})
describes $t_\mu $ in the $XY$ regime. On the other hand, in this regime
only vortices contribute to the temperature renormalization since such a
renormalization owing to spin-waves is absent (the interaction of spin waves
in the $XY$ model is due to topological effects only). Thus for the
temperature renormalization we have the system of RG equations \cite
{Kosterlitz,Jose}, which in our notations can be written as 
\begin{mathletters}
\label{EqXY}
\begin{eqnarray}
\mu \frac{d(1/t_\mu )}{d\mu } &=&32\pi ^2y_\mu ^2  \label{EqXY1} \\
\mu \frac{dy_\mu }{d\mu } &=&-y_\mu (2-\frac 1{2t_\mu })  \label{EqXY2}
\end{eqnarray}
It should be noted that the coupling constant for the vortex system is not $%
t $ (as for spin waves), but $y=\exp (-E_0/T)$ where $E_0$ is the energy of
a vortex core. Therefore equations (\ref{EqXY}) are applicable for small
enough $\mu $. Let $\mu _1\ll \sqrt{1-\eta }$ be the scale where we pass to
scaling (\ref{EqXY}). Then the solution of Eqs. (\ref{EqXY}) for $t>t_{KT}$
reads 
\end{mathletters}
\begin{equation}
\frac 1{t_\mu }=4+2C_1\tan \left( C_1\ln \frac \mu {\mu _1}+C_2\right)
\label{tmXY}
\end{equation}
where 
\begin{eqnarray}
C_1 &=&\sqrt{(8\pi y_1t_1)^2-(4t_1-1)^2}/(2t_1)  \nonumber \\
\tan C_2 &=&\frac{1-4t_1}{\sqrt{(8\pi y_1t_1)^2-(4t_1-1)^2}}
\end{eqnarray}
and $t_1\equiv t_{\mu _1},$ $y_1\equiv y_{\mu _1}$ are determined by 
\begin{eqnarray}
\frac 1{t_1} &=&\frac 1t-\ln \frac{\mu _0}{\sqrt{1-\eta }}+2\ln \frac t{t_1}%
+\Phi (\mu _1)  \label{t1y1} \\
y_1 &=&\frac 1{4\pi }\left[ \frac \mu 2\frac{d\Phi (\mu )}{d\mu }\right]
_{\mu =\mu _1}^{1/2}  \nonumber
\end{eqnarray}
It should be stressed that even if the original model is quantum one, the
resulting $XY$ model is classical since $\mu _1\ll \sqrt{1-\eta }\ll L_\tau
^{-1}$ ($L_\tau =JS/T$ for FM case and $L_\tau =c/T$ for AF case is a
characteristic length for quantum effects) and at scales much larger than $%
L_\tau $ quantum and classical systems becomes indistinguishable. Thus all
the quantum effects are already taken into account at the scales $\mu \gg 
\sqrt{1-\eta },$ where the behavior of RG trajectories is Heisenberg one.

The Kosterlitz-Thouless temperature $T_{KT}$ is determined by the equation
of the separatrix line for Eqs.(\ref{EqXY}) 
\begin{equation}
8\pi y_1=1/t_1-4,\;t=t_{KT}
\end{equation}
This line separates the low- and high-temperature phases. For small enough $%
\mu $ we have $\Phi (\mu )\rightarrow $const$,$ $d\Phi (\mu )/d\mu
\rightarrow 0$ and we have for $t_{KT}=T_{KT}/(2\pi JS^2)$ (or $T_{KT}/(2\pi
\rho _s)$ in AF-case)
\begin{equation}
t_{KT}=\left[ \ln (\mu _0/\sqrt{1-\eta })+2\ln (2/t_{KT})+C\right] ^{-1}
\label{tkt}
\end{equation}
where $C=4-6\ln 2-\Phi (\mu \rightarrow 0)$ is an universal constant. This
result is identical with that for the Curie (Neel) temperature of an
easy-axis magnet\cite{OurRG}, except for the constant $C,$ which needs not
be the same as for the easy-axis case.

In the critical region above $t_{KT},$%
\begin{equation}
\frac 1{8\pi }(t_{KT}^{-1}-t_{}^{-1})\ll 1,  \label{crit}
\end{equation}
the expression for the correlation length obtained from (\ref{tmXY}) reads 
\begin{equation}
\xi =\frac 1{\mu _1}e^{-C_2/C_1}\simeq \frac 1{\sqrt{1-\eta }}\exp \left( 
\frac A{2\sqrt{t_{KT}^{-1}-t_{}^{-1}}}\right)
\end{equation}
and has the same form as for the $XY$ model ($A$ is a constant). Under the
condition, opposite to (\ref{crit}), we have the standard Heisenberg
behavior \cite{Chakraverty} 
\begin{equation}
\xi =(C_\xi /\mu _0)t\exp (1/t)
\end{equation}

In the presence of interlayer coupling, the magnetic ordering at low enough
temperatures occurs. Due to topological effects, the transition temperature
grows up from $T_{KT}$, and not from zero. Note that in this case $T_{KT}$
plays a role of a crossover temperature from $2D$ to $3D$ $XY$ behavior
rather than a critical temperature, and the only true phase transition is
connected with the magnetic ordering at $T_C(T_N).$

In the case $\alpha \ll 1-\eta $ we choose $\mu _1$ such that $\alpha
^{1/2}\ll \mu _1\ll (1-\eta )^{1/2}.$ In terms of RG transformation (see
Fig.1), at $\mu =\mu _1$ we have not 2D, but quasi-2D $XY$ effective model
with the lattice constant $\mu _0/\mu _1$ and the interlayer coupling $(\mu
_0/\mu _1)^2\alpha _1$ where, as follows from (\ref{Eq4}), 
\begin{equation}
\alpha _1\equiv \alpha _{\mu _1}=\alpha t/t_1
\end{equation}
With further flow of the RG transformation we should arrive at the 3D $XY$
model. However, this part of the RG transformation meets with difficulties
owing to a complicated geometry of vortex loops (see Ref. \cite
{Chattopadhyay} and references therein). Instead of direct calculation of RG
trajectories, we use the same scaling arguments as in Ref. \cite{Hikami}.
The transition temperature can be estimated from the requirement that the
correlation length of the model without interlayer coupling ($\alpha =0$)
coincides with the characteristic scale of the crossover from 2D to 3D $XY$
model, $1/\alpha _1^{1/2}$ (in the units of the lattice constant of original
lattice). Then we have for the critical temperature $t_c=T_C/(2\pi JS^2)$
(or $T_N/(2\pi \rho _s)$) in the case $\alpha \ll 1-\eta $%
\begin{equation}
t_c=\left\{ \ln \frac{\mu _0}{\sqrt{1-\eta }}+2\ln \frac 2{t_{KT}}+C-\frac{%
A^2}{\ln ^2[(1-\eta )/\alpha ]}\right\} ^{-1}  \label{tc}
\end{equation}
The last term in the denominator determines the difference between $t_c$ and 
$t_{KT}$. Since this term can be not too small, we do not expand (\ref{tc})
in it.

The result (\ref{tc}) is qualitatively valid up to $\alpha $ of order of $%
1-\eta $ (in this case last term in the denominator leads to renormalization
of $C$ only). Consider now briefly the case $\alpha \gg 1-\eta .$ Then the
corrections to the RG result for the quasi-2D magnets\cite{OurRG} owing to
the easy-plane anisotropy are given by 
\begin{equation}
t_c=\left[ \ln \frac{\mu _0}{\sqrt{\alpha }}+2\ln \frac 2{t_c}+C^{\prime }+%
{\cal O}\left( \frac{(1-\eta )^{1/\psi }}{\alpha ^{1/\psi }}\right) \right]
^{-1}  \label{tc1}
\end{equation}
where $\psi =\nu _3(2-\gamma _\eta )$ is the crossover exponent, $\nu _3$ is
the corresponding critical exponent for the 3D Heisenberg model and $\gamma
_\eta $ is the anomalous dimensionality of the anisotropy parameter near 3D
Heisenberg fixed point, see, e.g., Ref. \cite{Amit}. The $\varepsilon $%
-expansion in the anisotropic $4-\varepsilon $ dimensional $\phi ^4$ model
for $\varepsilon =1$ (which has the same symmetry as the model under
consideration) yields $\psi \simeq 0.83,$ see Ref. \cite{Amit}. For an
antiferromagnet, the constant $C^{\prime }\simeq -0.066$ was calculated
within the $1/N$ expansion\cite{Our1/N}. Unlike (\ref{tc}), the last term in
the denominator of (\ref{tc1}) has not inverse-logarithmic form. This is a
consequence of the fact that the correlation length in the 3D Heisenberg
model does not demonstrate the exponential behavior ($\nu _3$ is finite). By
this reason the correction in the denominator of (\ref{tc1}) is small and
can be neglected.

Finally, we consider the experimental situation for layered magnets. The
mostly investigated easy-plane system is the compound K$_2$CuF$_4.$ This is
a $S=1/2$ ferromagnet with $T_{KT}=5.5$K,$\;T_C=6.25$K and the parameters $%
J=20$K$,$ $1-\eta =0.04,\;\alpha =6\cdot 10^{-4}$ (see, e.g., Ref. \cite
{Joungh}). Substituting these values into (\ref{tkt}) and (\ref{tc}) we
obtain $C\simeq -0.5$ and $A\simeq 3.5.$ Note that the formula (\ref{TKTs})
yields the value $T_{KT}=11.4$K which is much larger than the experimental
one.

Another example of a quasi-2D FM $XY$-like system is the stage-2 NiCl$_2$
graphite interlayer compound with $S=1$. According to Ref. \cite{Joungh}, $%
J=20K$, $1-\eta =8\cdot 10^{-3}\ $and $\alpha =5\cdot 10^{-5}$. Using the
same values of $A$ and $C$ as for K$_2$CuF$_4,$ we calculate $T_{KT}=17.4$ K
and $T_C=18.7$ K, which is in agreement with experimental data (both values $%
T_{KT}$ and $T_C$ lie in the region $18-20$ K). At the same time, using the
formula (\ref{TKTs}) yields $T_{KT}=35.3$K, which is again twice larger as
compared to the experimental value.

We have also applied our results to the compound BaNi$_2$(PO$_4$)$_2$ which
is a $S=1$ antiferromagnet with $|\,J\,|=22.0$ K and easy-plane anisotropy $%
1-\eta =0.05,\;\alpha =1\cdot 10^{-4},$ see Ref. \cite{Joungh}. We obtain $%
T_{KT}=23.0$ K which coincides with the experimental value and $T_N=24.3$ K,
again in excellent agreement with $T_N^{\exp }=24.5\pm 1$K. Note that in
spite of $T_{KT}\sim |J|S$ for this compound, the true criterium of the
quantum regime is $(T/JS)^2\ll 32$ (see Ref. \cite{OurRG}), and this case
also should be considered as quantum one.

To conclude, we have investigated the Heisenberg model with a weak easy-axis
anisotropy. We have performed the two-loop RG transformation with unknown
function $\Phi (\mu )$ (which takes into account the contribution of higher
loops and non-spin-wave excitations) and joined the results with well-known
behavior of the RG trajectories in the 2D $XY$ model. In such a way we have
obtained simple analytical expressions for the Kosterlitz-Thouless and Curie
(N\'eel) temperatures. These expressions contain two constants $A,C$ which
are still indeterminate within the RG approach. The calculation of these
constants, as well as of the corresponding parameters for the isotropic
quasi-2D and easy-axis 2D Heisenberg model \cite{OurRG}, is possible by
numerical (e.g., by the quantum Monte-Carlo) methods\cite{Prim}. At the same
time, our results already enable one to estimate the Kosterlitz-Thouless
temperature (and also to determine the difference $T_C(T_N)-T_{KT}$) with
the accuracy which is sufficient to fit experimental data on layered
magnets, unlike the simplest expression (\ref{TKTs}).

We are grateful to B.N.Shalaev for useful discussions.

{\sc Figure caption}

Schematic picture of the RG\ trajectories in layered magnets. Left-hand
side: the flow from the 2D easy-axis Heisenberg (H+EA) to 2D Ising model.
Right-hand side: the flow from the 2D easy-plane Heisenberg (H+EP) to 2D $XY$%
\ model. The inflection points $c_1$, $c_2$ mark the crossover regions. The
dashed lines are for the corresponding quasi-2D models.


\begin{references}
\bibitem{Prim}  In the AF case one can use also the $1/N$-expansion.
However, unlike the isotropic quasi-2D case\cite{Our1/N}, in the presence of
anisotropy this expansion becomes inapplicable in the critical region (which
gives a considerable contribution to $C$) because of small values $N=1$ and $%
N=2$ for the Ising and $XY$ models.
\end{references}
\end{document}